\documentclass[a4paper,twoside]{article}
\newcommand{\affil}[1]{$^{\rm #1}$}
\usepackage[authoryear]{natbib}
\bibpunct{(}{)}{;}{a}{}{,}
\usepackage{graphicx}
\date{\small  Received 2009 March 10, accepted 2009 August 27} 
\newcommand{\kms}{\mbox{km\,s$^{-1}$}}
\title{\large\bf\flushleft A Survey of IUE Spectra of the Active Binary System UX Arietis}
\author{\parbox{\textwidth}{\flushleft
\vspace{-0.5cm}
{\it Fehmi Ekmek\c{c}i\affil{A}}\\
\vspace{0.4cm}
{\small \affil{A}\ Ankara University, Faculty of Science, Department of
Astronomy and Space Sciences, 06100 Tando\u{g}an, Ankara, Turkey.}
{\small Email: fekmekci@science.ankara.edu.tr}}}
\begin{document}
\twocolumn[
\begin{changemargin}{.8cm}{.5cm}
\begin{minipage}{.9\textwidth}
\vspace{-1cm}
\maketitle
%
%
\small{\bf Abstract: To investigate the ultraviolet (UV) activity of the bright, non-eclipsing, double-lined 
spectroscopic binary system {\rm UX Ari}, IUE spectra (194 images) were taken from the IUE archive. The spectra, obtained during the period 1978-1996, show emission lines originating in the chromosphere and transition region. The long-wavelength low dispersion spectra were examined for ultraviolet excess by comparing the UV continuum level of {\rm UX Ari} with the levels of {\rm $\kappa$  Cet}  and {\rm $\eta$  Cep} in the spectral range 2100 $\AA$ - 3200 $\AA$ . The individual MgII h and k emission-line fluxes of component stars show that the contributions to the activity of the system for G5 V and K0 IV are about 20\%  and  80\%, respectively. Apart from the flare event observed on 1979 Jan 1, there are some flux enhancements in the years 1987, 1991 and 1994 which may suggest existence of a periodicity of about 7-9 years. Examination of the 
ultraviolet excess in the system showed that there is some UV excess in {\rm UX Ari}, which varies from 1\% up to 24\% with the exception of two images which showed no UV excess. The results revealed that there was an agreement on the source of emission regions which could be attributed to the magnetic activity. The contribution of G5 V and K0 IV components to the MgII activity of the system suggests a need to take into consideration the spot distribution not only on the surface of K0 IV but also on the surface of G5 V component of {\rm UX Ari}.}


\medskip
\medskip
\end{minipage}
\end{changemargin}
]
\small

\section{Introduction}

{\rm UX Ari} is a bright non-eclipsing double-lined spectroscopic binary (${\it P}_{orb} = 6.43791$ days) of 
spectral type G5 V + K0 IV \citep{carlos71}. The primary star, in this case the K0 IV component, shows 
chromospheric activity and is responsible for the majority of the activity shown by the system, 
as in most such {\rm RS CVn} systems. Some photometric and orbital characteristics of {\rm UX Ari} are 
summarized in Table 1.  

\citet{huen89} summarized the previous observations that had been done by numerous investigators, including photometric, spectroscopic, X-ray, radio and ultraviolet features. They suggested that the excess absorption was due to the mass-transfer activity resulting from the Roche lobe over-flow of the K star and accretion onto G star by analyzing their $H_{\alpha}$ and $H_{\beta}$ observations obtained as fibre-optic, echelle, CCD spectra. \citet{vogt91} derived an accurate measurement of differential rotation with the opposite feature to that of the Sun using Doppler Imaging Technique. They also showed the spot distribution, which was quite complex, and the primary component, which had a large, stable polar spot.

        \begin{table}[ht!]
        \small
        \flushleft
        \caption{The characteristics of {\rm UX Ari}}
        \label{tab1}
	\medskip
        \begin{flushleft}
        \begin{tabular}{lll}
        \hline
        Parameter         &                                & References \\
	\hline
        $\alpha_{2000}$ &03$^{\rm h}$26$^{\rm m}$ 35$^{\rm s}$.36  & a \\
        $\delta_{2000}$ &+28$^{\rm o}$ 42$^{\rm '}$ 55$^{\rm "}$.2 & a \\
        Distance [pc]     &50.23                           & a \\
	${\it P}_{orb.}$  &6$^{\rm d}$.43791               & b \\
	Sp.Type           & hot : G5 V                     & c \\
	                  & cool : K0 IV                   &       \\
        Masses/M$_{\odot}$& hot  : $\geq 0.93$            & c\\
                          & cool : $\geq 0.71$             &                         \\
        Radii/R$_{\odot}$             & hot  : 0.93       & c \\
                          & cool : $> 4.7$                 &                         \\
        M$_{v}$           & 2$^{\rm m}$.5                  & c \\
        {\it V}           & 6$^{\rm m}$.38                 & c \\
        {\it B-V}         & 0$^{\rm m}$.91                 & c \\
        {\it U-B}         & 0$^{\rm m}$.48                 & c \\
        {\it i}           & 60$^{\rm o}$                   & c \\
        \hline
        \end{tabular}
        \end{flushleft}
        $^{(a)}${The Hipparcos and Tycho Catalogue.}\\
        $^{(b)}${Carlos \& Popper(1971).}\\
        $^{(c)}${Strassmeier et al.(1993).}
        \end{table}

\citet{duem01} attempted to improve the orbital measurements of {\rm UX Ari} by using the published radial velocities together with their high-accuracy data. They improved the set of orbital parameters and found that the $\gamma$ velocity of the system has a systematic variation with time. They concluded that {\rm UX Ari} seemed to be a triple system. The excess emission/absorption in some chromospheric lines (see e.g. Montes et al. 1995a,b for $H_{\alpha}$ emission, and Huenemoerder et al. 1989 for $H_{\alpha}$ absorption) or the excess in continuum levels of various spectral ranges, if they exist, could provide evidence for the source of activity being dependent on rotation (related to magnetic activity), or an accretion stream from Roche-lobe overflow of the primary, or some process that occurs in hot circumstellar gas \citep[see][]{rhomb77}, 
respectively. Therefore, the examination of the ultraviolet excess would also be advantageous.

\citet{ekm93} studied 65 IUE spectra observed in the 1978-1987 period(with  IUESIPS  reduction). It was shown that emission-line fluxes vary with the orbital phase and that the dependence of the line fluxes on orbital phase was well correlated with the photometric light variation. This  correlation  might indicate more active  chromospheric  regions above the photospheric spot regions. \citet{ekm93} also measured fluxes of the
individual IUESIPS emission lines of the component stars of {\rm UX Ari} and calculated that the contributions to the activity of the system for G5 and K0 were about 1/4 and 3/4  respectively. Another characteristic of the IUE spectra (with IUESIPS reduction) was an absorption feature observed on the peak of the IUSIPS k profiles of the K0 IV component, which was observed to shift together with the emission profile as the star revolved in its orbit \citep{ekm93}. Based on this absorption feature, it is suggested that the circumstellar matter around the K0 IV component may be responsible for this absorption. 

\citet{nichols96} summarized a new calibration together with new image processing techniques. They pointed out that the wavelength and absolute flux errors in the IUESIPS processing could be corrected by NEWSIPS progressing and that it would be better to re-analyze all the IUE spectra with NEWSIPS reduction. With this aim, in the present paper all 194 images of IUE-NEWSIPS spectra of {\rm UX Ari} observed in 1978-1996 period have been analyzed to check the validity of the previous findings of \citet{ekm93}. This paper shows that all the integrated emission-line fluxes of short-wavelength low-dispersion spectra have a variation with time and orbital phase, but that the variation with time was not as clear as that with the orbital phase. Examination of the ultraviolet excess shows that some UV excess in {\rm UX Ari} exists and varies from 1\% up to 24\% in time. Comparison of the MgII radial velocities with those of visible spectra (Carlos \& Popper 1971;Duemmler \& Aarum 2001) showed that the scattering in the UV data is likely to come from chromospheric activity caused by a magnetic dynamo that produced loops in active region.
 
\section{IUE data and spectral analysis}

The IUE spectra of {\rm UX Ari} have been taken from the NASA IUE archieve using the IDL (Interactive Data Language) Program. All the spectra have undergone NEWSIPS reduction. The spectra consist of 22 LWP, 2 LWR and 86 SWP images in low dispersion, and 69 LWP, 12 LWR and 3 SWP images in high resolution. The log of IUE images are given in Table 2. The images studied by other authors in the past were denoted by asteriks in the 'Comment' column of Table 2. IUE obtained spectra at both low (6  \AA  resolution) and high dispersion ($\lambda / \delta\lambda  \sim  10000$), with the short-wavelength prime (SWP, 1151 - 2000 \AA), long wavelength prime (LWP, 1850 - 3400 \AA), and long-wavelength redundant (LWR, 1850 - 3400 \AA) cameras \citep{nichols96}.
 
The spectra show emission lines originating in the chromosphere and transition region. The flux in a given line was obtained by computing the area contained in the spectral region above the continuum or background levels near the wings of the line. The emission line fluxes were computed based on Gaussian profile-fitting procedures. The fitting procedures were made by means of the CURFIT program of \citet{beving69}. Some results of the Gaussian fits are shown in Figure 1 for low dispersion image, SWP02375, and in Figure 6 for high-resolution images, LWP14085 and LWP14130, in the range that includes the Mg II h and k lines. The overall shapes of the line profiles can be reasonably well matched by 1 or 2 Gaussian components for low dispersion SWP spectra and by 3 Gaussian components for Mg II h and k profiles in high-resolution images. In the case of Si IV or Si II lines there are two Gaussian components for the multiplets in the fitting procedure. The strong emission in these lines originates from the K0 IV star rather than G5 V star of {\rm UX Ari} system. In the case of the Mg II h and k profiles, a Gaussian profile has been attributed to each component of the system in the fitting procedures. A third Gaussian absorption profile represents the interstellar absorption component 
(Figure 6). In the Gaussian fitting procedures, since the effect of interstellar absorption can be removed 
and this removal does not show any significant effect in comparison with the observational errors on flux and wavelength of the IUE images, the uncertainties in the strength and velocity of Mg II h and k lines can be estimated independently based on the observational errors. Therefore, in this analysis we can be confident that there were no effects of interstellar absorption on determination of the secondary's contribution to the 
activity of the system. The orbital phases that correspond to the mid-time of IUE observations were computed with the ephemeris

       \begin{equation}
        \rm HJD = (2440133.766 + 6.43791\times{\it E})days,
       \end{equation}
for which the zero phase corresponds to conjunction with the primary (K0 IV) component in 
front \citep{carlos71}.

        \begin{table*}
        \scriptsize
        \caption{The log of IUE observations of UX Ari}
        \smallskip
        \begin{center}
        \begin{tabular}{llllllllll}
        \hline
        Image&Disp& HJD(mid)&Exp.Time&Comment&Image&Disp& HJD(mid)&Exp.Time&Comment\\
             &    &         & (sec.) &   &     &    &    & (sec.) &   \\
        \hline
        LWR02081& H& 2443736.0752&   720&        *        &LWP11752& L& 2447068.3635&    90&      \\
        SWP02301& L& 2443736.1016&  2700&        *        &LWP11753& L& 2447068.4835&    90&      \\
        LWR02082& H& 2443736.1354&  1800&        *        &LWP11754& L& 2447068.6115&    90&      \\
        LWR02111& H& 2443739.8502&  1800&        *        &LWP11755& L& 2447068.7315&    90&      \\
        SWP02336& L& 2443739.8983&  5400&        *        &LWP11756& H& 2447068.7764&  3000&      \\
        LWR02136& H& 2443741.9434&  1800&        *        &LWP11757& H& 2447068.8277&  1500&      \\
        SWP02351& L& 2443741.9853&  4200&        *        &LWP11758& H& 2447068.8727&  1500&      \\
        LWR02158& H& 2443743.9244&  1800&        *        &LWP11760& L& 2447068.9725&    90&      \\
        SWP02375& L& 2443743.9643&  4200&        *        &LWP11761& L& 2447068.9975&    90&      \\
        LWR03344& H& 2443874.5894&  1800&        *        &LWP11762& H& 2447069.0424&  3000&      \\
        SWP03766& L& 2443874.6273&  4200&        *        &LWP11763& L& 2447069.0845&    90&      \\
        SWP03855& L& 2443882.7704&  1800&        *        &LWP11764& L& 2447069.2335&    90&      \\
        LWR03432& H& 2443882.7963&  1080&        *        &LWP11765& L& 2447069.3595&    90&      \\
        LWR06261& H& 2444207.2292&   900&                 &LWP11766& L& 2447069.4825&    90&      \\
        SWP07267& L& 2444207.2668&  4800&        *        &LWP11767& L& 2447069.5875&    90&      \\
        LWR06329S& L& 2444215.9747&  120&                 &LWP11768& L& 2447069.6135&    90&      \\
        LWR06329L& L& 2444215.9784&  240&                 &LWP11769& L& 2447069.6375&    90&      \\
        SWP07342& L& 2444216.0098&  4800&       *         &LWP11770& L& 2447069.6635&    90&      \\
        LWR06330& H& 2444216.0524&  1800&                 &LWP11771& H& 2447069.7233&  5400&      \\
        SWP07423& L& 2444225.4259& 12600&        *        &LWP14051& H& 2447419.8604&  3000&      \\
        LWR10244& H& 2444693.3949&  1200&        *        &LWP14052& H& 2447419.9127&  1500&      \\
        SWP13612& L& 2444693.4234&  3000&        *        &LWP14053& H& 2447419.9472&   720&      \\
        SWP15211& H& 2444886.6123& 27000&*,noisy,excluded&LWP14084& H& 2447422.9887&  1500&       \\
        LWR11729& H& 2444886.7817&  1500&        *        &LWP14085& H& 2447423.0317&  1500&      \\
        SWP15240& H& 2444889.5829& 24000&*,noisy,excluded&LWP14086& H& 2447423.0926&  4080&       \\
        LWR11756& H& 2444889.7319&  1200&                 &LWP14130& H& 2447432.8294&  3000&      \\
        SWP26730& L& 2446334.6765&   600&        *        &LWP14131& H& 2447432.8817&  1500&      \\
        SWP26730& L& 2446334.6875&   600&        *        &LWP14132& H& 2447432.9334&  3000&      \\
	    SWP26730& L& 2446334.7095&   600&        *        &LWP14152& H& 2447435.9527&  1500&      \\
        LWP06815& H& 2446334.7232&   900&                 &LWP14153& H& 2447435.9967&  1500&      \\
        SWP26731& L& 2446334.7415&   600&        *        &LWP14220& H& 2447448.7844&  3000&      \\
	    SWP26731& L& 2446334.7555&   600&        *        &LWP14221& H& 2447448.8367&  1500&      \\
	    SWP26731& L& 2446334.7675&   600&        *        &LWP14222& H& 2447448.8777&  1500&      \\
        LWP06816& H& 2446334.7892&   900&                 &LWP14263& H& 2447451.9177&  1500&      \\
	    SWP26732& L& 2446334.8095&   600&        *        &LWP14264& H& 2447451.9587&  1500&      \\
	    SWP26732& L& 2446334.8225&   600&        *        &LWP14265& H& 2447452.0141&  3480&      \\
        SWP26732& L& 2446334.8345&   600&        *        &LWP18569& H& 2448116.5273&  1080&      \\
        LWP06817& H& 2446334.8552&   900&                 &SWP39449& L& 2448116.5547&  1500&      \\
	    SWP26733& L& 2446334.8755&   600&*,noisy,excluded&SWP39460& L& 2448118.0615&  1806&       \\
	    SWP26733& L& 2446334.8925&   600&*,noisy,excluded&LWP18584& H& 2448118.0829&  1200&       \\
	    SWP26733& L& 2446334.9035&   600&*,noisy,excluded&SWP39470& L& 2448118.9079&  2400&       \\
        LWP06818& H& 2446334.9252&   900&                 &LWP18597& H& 2448119.8816&  1320&      \\
	    SWP26734& L& 2446334.9435&   600& noisy, excluded&LWP18607& H& 2448121.0516&  1320&       \\
	    SWP26734& L& 2446334.9575&   600& noisy, excluded&SWP39476& L& 2448121.0769&  2400&       \\
	    SWP26734& L& 2446334.9658&   144& noisy, excluded&SWP42405& L& 2448505.9212&  2100&       \\
	    LWP06819& H& 2446334.9952&   900&                 &LWP21171& H& 2448505.9433&  1080&      \\
	    SWP26735& L& 2446335.0145&   600& noisy, excluded&SWP42416& L& 2448507.9249&  2400&       \\
	    SWP26735& L& 2446335.0265&   600& noisy, excluded&LWP21187& H& 2448507.9485&   600&       \\
        LWP09864& H& 2446801.4872&   900&                 &LWP21200& H& 2448508.9073&  1080&      \\
	    SWP30026& L& 2446801.5055&   600&        *        &SWP42427& L& 2448508.9379&  2400&      \\
	    SWP30026& L& 2446801.5165&   600&        *        &LWP21208& H& 2448509.8913&  1080&      \\
	    SWP30026& L& 2446801.5285&   600&        *        &SWP42435& L& 2448509.9209&  2400&      \\
        LWP09865& H& 2446801.5512&   900&        *        &LWP21222& H& 2448511.8890&  1380&      \\
	    SWP30027& L& 2446801.5725&   600&        *        &SWP42448& L& 2448511.9179&  2400&      \\
	    SWP30027& L& 2446801.5825&   600&        *        &LWP21236& H& 2448513.8990&  1380&      \\
	    SWP30027& L& 2446801.5955&   600&        *        &SWP42461& L& 2448513.9269&  2400&   *  \\
        LWP09866& H& 2446801.6172&   900&        *        &LWP28935& H& 2449585.1275&   600&      \\
        SWP30028& L& 2446801.6375&   600&        *        &SWP51857& L& 2449585.1502&   900&      \\
        SWP30028& L& 2446801.6485&   600&        *        &SWP51858& L& 2449585.1842&   900&      \\
        SWP30028& L& 2446801.6615&   600&        *        &LWP28940& H& 2449585.9735&   600&      \\
        LWP09867& H& 2446801.6852&   900&        *        &SWP51866& L& 2449585.9952&   900&      \\
        SWP30029& L& 2446801.7045&   600&        *        &SWP51867& L& 2449586.0272&   900&      \\
        SWP30029& L& 2446801.7165&   600&        *        &SWP51872& L& 2449586.9082&   900&      \\
        SWP30029& L& 2446801.7275&   600&        *        &LWP28943& H& 2449586.9205&   600&      \\
        LWP09868& H& 2446801.7502&   900&        *        &SWP51873& L& 2449586.9432&   900&      \\
        SWP30030& L& 2446801.7705&   600&        *        &SWP51884& L& 2449587.9022&   900&      \\
        SWP30030& L& 2446801.7805&   600&        *        &LWP28950& H& 2449587.9145&   600&      \\
        LWP11745& H& 2447067.8534&  3000&                 &SWP51885& L& 2449587.9392&   900&      \\
        LWP11746& H& 2447067.9037&  1500&                 &SWP51961& L& 2449592.8712&   900&      \\
        LWP11747& L& 2447067.9365&    90&                 &LWP29029& H& 2449592.8835&   600&      \\
        LWP11748& L& 2447067.9595&    90&                 &SWP51962& L& 2449592.9092&   900&      \\
        LWP11749& H& 2447068.0067&  1500&                 &SWP51975& L& 2449593.8772&   900&      \\
        LWP11750& H& 2447068.0673&  4200&                 &LWP29040& H& 2449593.8895&   600&      \\
        SWP31952& H& 2447068.6157& 89280& noisy, excluded&SWP51976& L& 2449593.9112&   900&       \\
        LWP11751& L& 2447068.2375&    90&                 &LWP29052& H& 2449594.9765&   600&      \\        	
	\hline
        \end{tabular}
        \end{center}
        \end{table*}

 
        \begin{table*}
        \scriptsize
        \addtocounter{table}{-1}
        \caption[]{( Continued )}
        \smallskip
        \begin{center}
        \begin{tabular}{llllllllll}
        \hline
        Image&Disp& HJD(mid)&Exp.Time&Comment&Image&Disp& HJD(mid)&Exp.Time&Comment \\
             &    &         & (sec.) &     &     &    &         & (sec.) &     \\
        \hline
        SWP51986& L& 2449594.9992&   900&        &SWP52056& L& 2449602.9102&   900&      \\
        SWP51996& L& 2449597.0522&   900&        &LWP29118& H& 2449602.9275&   600&      \\
        LWP29070& H& 2449597.0695&   600&        &SWP52057& L& 2449602.9492&   900&      \\
        SWP51997& L& 2449597.0922&   900&        &LWP29127& H& 2449604.0545&   600&      \\        
        LWP29071& H& 2449597.1125&   600&        &SWP52063& L& 2449604.0772&   900&      \\
        LWP29077& H& 2449598.0515&   600&        &LWP29128& H& 2449604.1025&   600&      \\
        SWP52007& L& 2449598.0742&   900&        &LWP29137& H& 2449605.0615&   600&      \\
        LWP29078& H& 2449598.0945&   600&        &SWP52070& L& 2449605.0872&   900&      \\
        SWP52008& L& 2449598.1135&   600&        &LWP29138& H& 2449605.1145&   600&      \\
        SWP52016& L& 2449599.0652&   900&        &SWP52078& L& 2449606.0732&   900&      \\
        LWP29085& H& 2449599.0765&   600&        &LWP29142& H& 2449606.0925&   600&      \\
        SWP52017& L& 2449599.0992&   900&        &LWP29149& H& 2449606.8835&   600&      \\
        SWP52023& L& 2449600.0622&   900&        &SWP52086& L& 2449606.9052&   900& noisy, excluded\\
        LWP29091& H& 2449600.0805&   600&        &SWP52087& L& 2449606.9342&   900& noisy, excluded\\
        SWP52024& L& 2449600.1012&   900&        &LWP31888& L& 2450100.3785&    90&      \\
	    SWP52034& L& 2449601.0572&   900&        &SWP56587& L& 2450100.4143&  5400&   *  \\
        LWP29099& H& 2449601.0735&   600&        &LWP31894& L& 2450103.3545&    90&      \\
        SWP52035& L& 2449601.0942&   900&        &SWP56624& L& 2450103.3913&  5400&      \\
        LWP29100& H& 2449601.1135&   600&        &LWP31895& L& 2450103.3885&    90&      \\
        SWP52046& L& 2449602.0722&   900&        &LWP31896& L& 2450103.4245&    90&      \\
        LWP29109& H& 2449602.0895&   600&        &SWP56630& L& 2450105.3971&  4680&      \\
        SWP52047& L& 2449602.1092&   900&        &LWP31903& L& 2450105.3995&    90&      \\                                          
        \hline 
        \end{tabular}
        \end{center}
        $^{*}${Images studied by other authors(see http://archieve.stsci.edu/iue/search.php) in the past.}\\
	    \end{table*}

\subsection{Short-wavelength, Low - Dispersion Spectra}

The most prominent feature seen in the spectra is $Ly\alpha$  profile. Since this line is blended with geocoronal $Ly\alpha$  and overexposed throughout the observations, it is not included in the line analysis. Due to  6-$\AA$  resolution, some of the emission lines are unresolved or partially resolved multiplets. The 
identified emission lines of SWP spectra (see Figure 1) are OI ($\lambda$ $\lambda$1302,1305), CI ($\lambda$1657), SiII ($\lambda$ $\lambda$1808,1817), CII ($\lambda$ $\lambda$1334, 1335), HeII ($\lambda$1639), NV ($\lambda$1238), SiIV ($\lambda$ $\lambda$1393, 1402), and CIV ($\lambda$ $\lambda$1548, 1550). Figures 2 to 4 show the integrated emission Carbon line fluxes of low-dispersion spectra as a function of time (orbital epoch) and orbital phase. These lines originate in chromosphere and transition region. The same trend was seen for the other line fluxes (the lines mentioned above).  


\begin{figure*}[h!t]
\centering
\vspace{0.1cm}
\includegraphics[angle=0, width=0.75\textwidth]{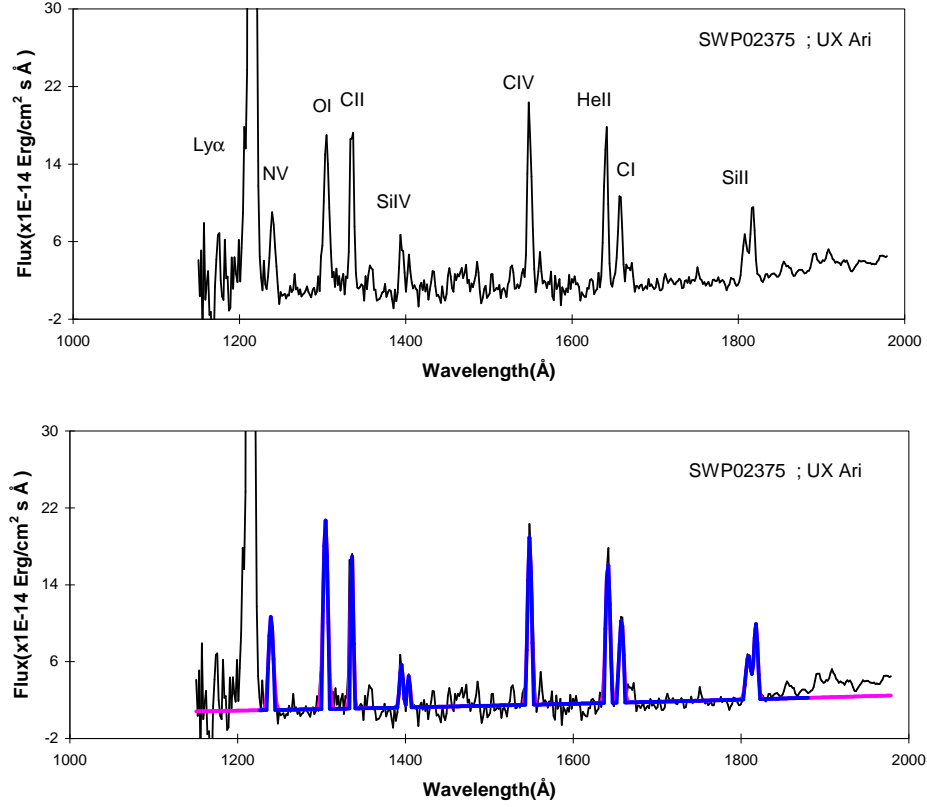}
\caption{The Gaussian fits to chromospheric and transition-region lines of {\rm UX Ari}
          appeared in the SWP02375 low-dispersion image.}
\label{fig1}
\end{figure*}

\begin{figure*}[h!t]
\vspace{0.1cm}
\begin{center}
\includegraphics[width=0.65\textwidth]{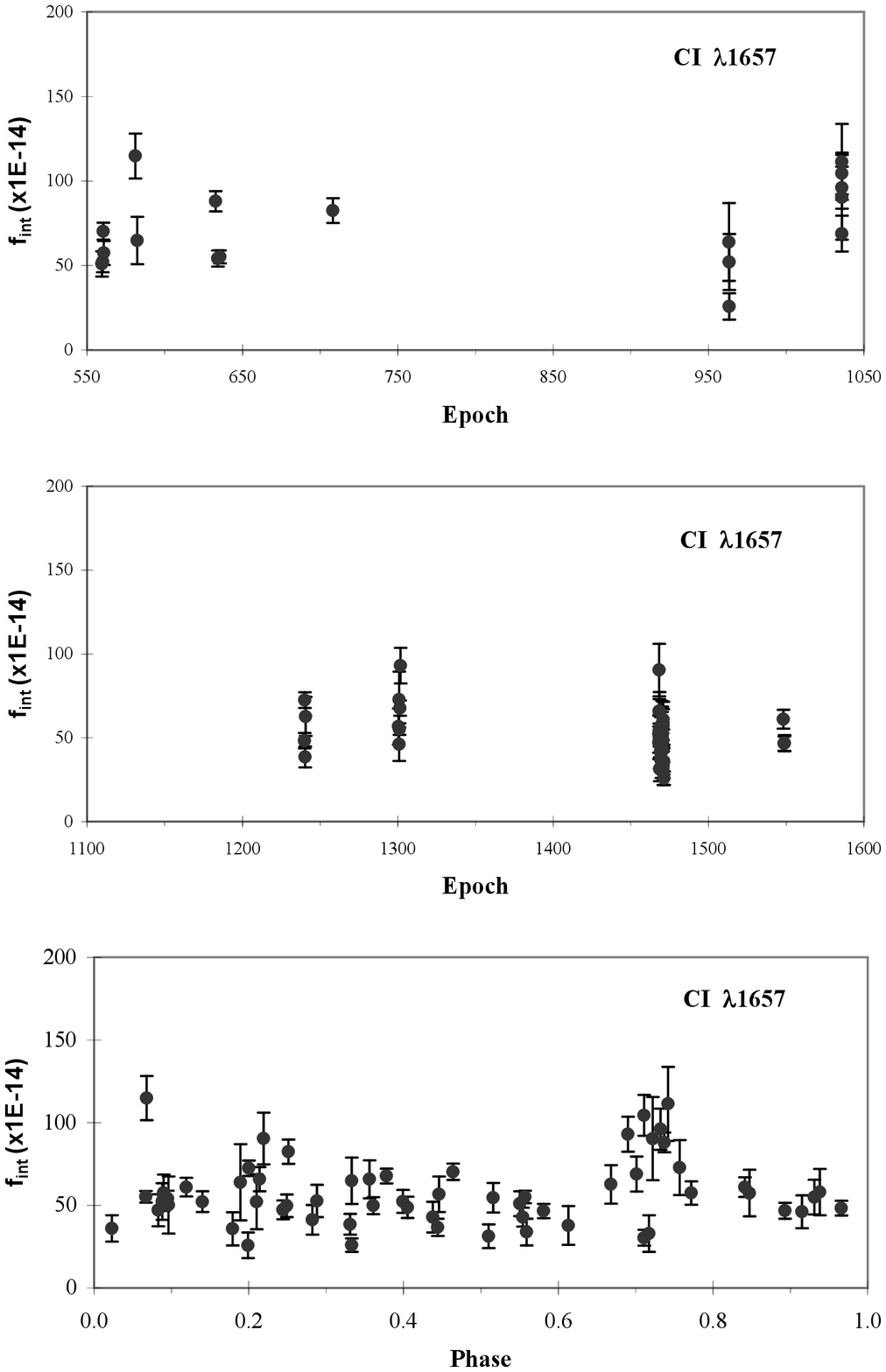}
\caption{Low dispersion CI emission line (originates in the middle chromosphere)
          fluxes as a function of time and orbital phase. The line fluxes are in
          units of ergs  $cm^{-2}$  $s^{-1}$.}
\label{fig2}
\end{center}
\end{figure*}

There is a flare event near the orbital phase $\sim 0.07$ (SWP03766). Apart from this flare event there are two rises in flux at phases $\sim 0.20$ and $\sim 0.70$ (Figures 2-4). It can be clearly seen that there are variations in the chromospheric and transition-region line fluxes with time and orbital phase. Since some of these low dispersion spectra (five images) are very noisy and have indeterminate lines, they were excluded from the analysis. These images are SWP26733, SWP26734, SWP26735, SWP52086 and SWP52087. The scattering of the emission line fluxes of the images, taken at the same epoch (see Figures 2-4) arose from the variation of flux with orbital phase.

\begin{figure*}[h!t]
\vspace{0.1cm}
\begin{center}
\includegraphics[width=0.65\textwidth]{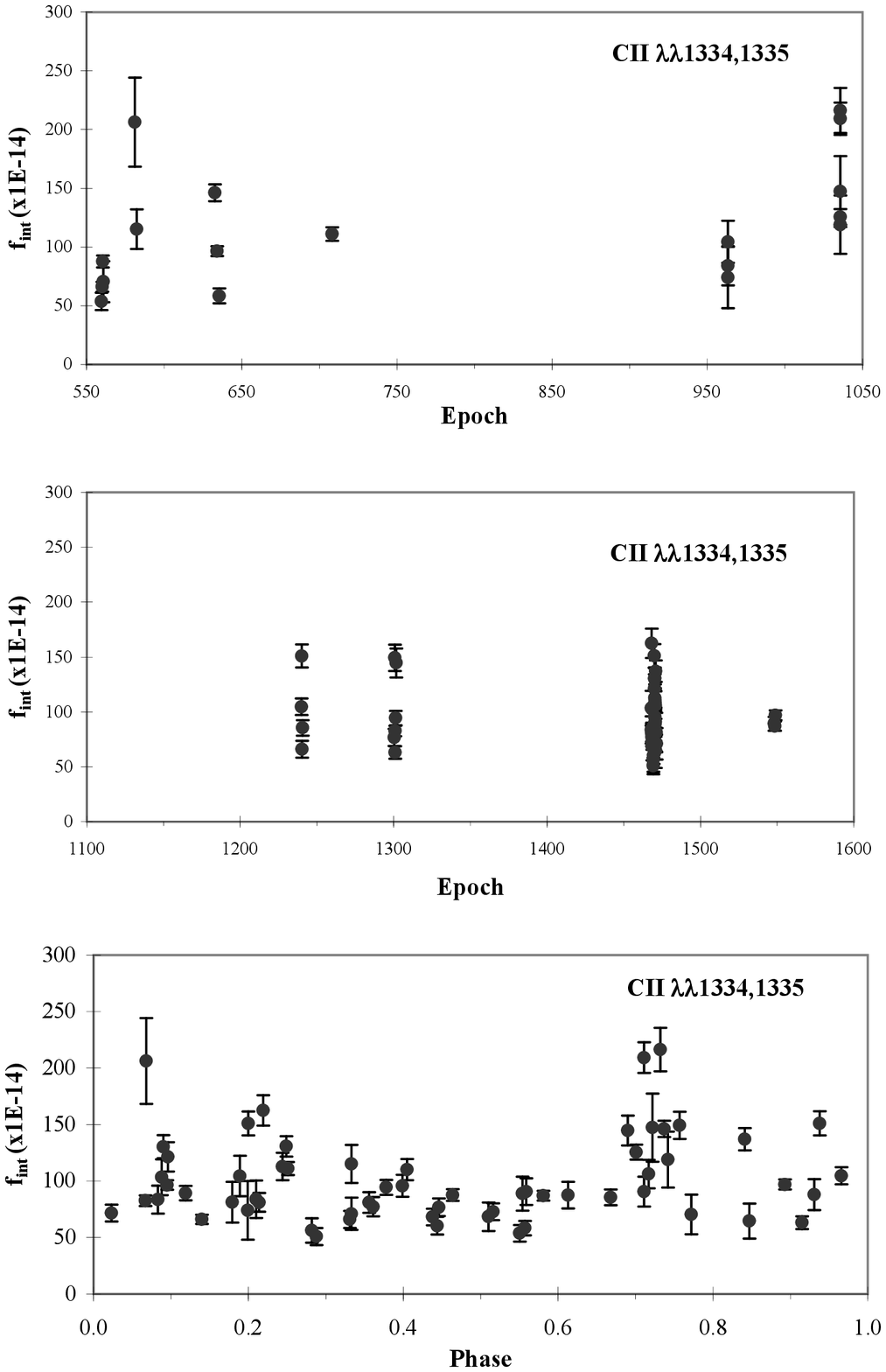}
\caption{Low dispersion CII emission line (originates in the upper chromosphere)
          fluxes as a function of time and orbital phase. The line fluxes are in
          units of ergs  $cm^{-2}$  $s^{-1}$.}
\label{fig3}
\end{center}
\end{figure*}

\subsection{Short-Wavelength, High - Dispersion Spectra}

There exist only three images of {\rm UX Ari} system taken by the SWP camera in high resolution: SWP15211, SWP15240 and SWP31952. NV ($\lambda$1239), OI ($\lambda$ $\lambda$ 1302, 1305), CII ($\lambda$ $\lambda$ 1334, 1335), SiIV ($\lambda$ $\lambda$ 1394, 1403), CIV ($\lambda$ $\lambda$ 1548, 1550), HeII ($\lambda$1640), CI ($\lambda$1657), SiII ($\lambda$ $\lambda$ 1808, 1817), SiIII ($\lambda$1892),  and CIII ($\lambda$1909) lines were looked for in these three images to evaluate the fluxes of these lines together with those of the low dispersion spectra. Unfortunately, all three images have a lot of reseau (in the ITF), permenant ITF artifact, saturated pixels, warning tracks, RAW - SCREEN cosmic rays/bright spots, positively extrapolated ITF, and very noisy data in the range of these lines. Hence, these lines were hardly seen in the images SWP15211, SWP15240, and SWP31952 which were taken at orbital epochs 738.259, 738.721 and 1077.190, respectively. Therefore, these three images were excluded from the analysis.

\begin{figure*}[h!t]
\vspace{0.1cm}
\begin{center}
\includegraphics[width=0.65\textwidth]{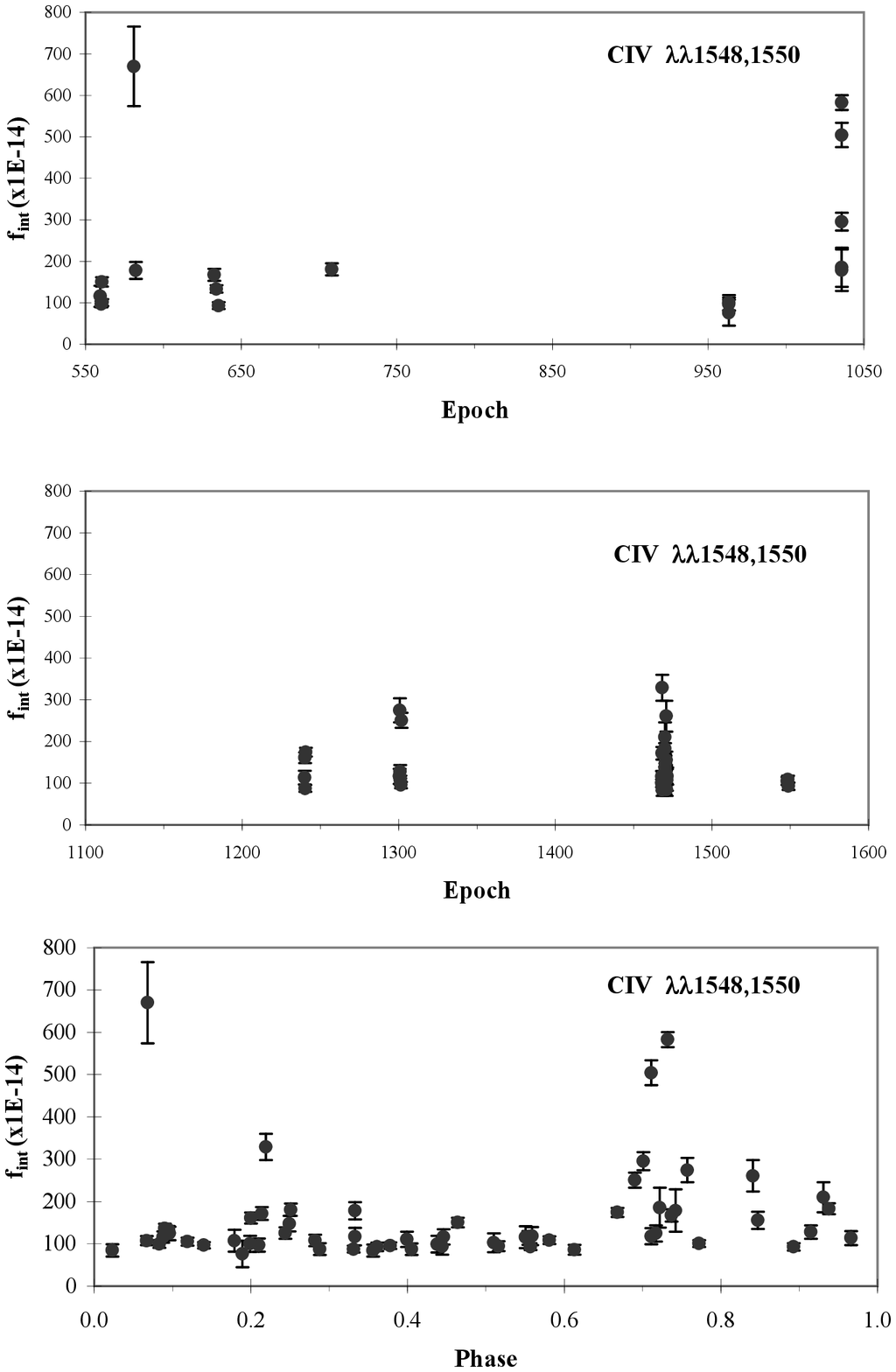}
\caption{Low dispersion CIV emission line (originates in the transition region)
          fluxes as a function of time and orbital phase. The line fluxes are in
          units of ergs  $cm^{-2}$  $s^{-1}$.}
\label{fig4}
\end{center}
\end{figure*}
	
\subsection{Long-Wavelength, Low - Dispersion Spectra}

The long-wavelength low dispersion spectra (24 images) were examined for ultraviolet excess, by comparing the ultraviolet continuum level of {\rm UX Ari} (G5 V + K0 IV) with the levels of {\rm $\kappa$  Cet} (G5 V) and {\rm $\eta$  Cep} (K0 IV) in the same spectral range between $2100 \AA$ and $3200 \AA$. After polynomial fitting for the continuum levels (see Figure 5) and computing the integrated flux measured on Earth (f)between $2100 \AA$ and $3200 \AA$, the continuum level analysis can be carried out by converting the flux measured on Earth to the surface integrated flux (F) of a star by:

	\begin{equation}
         F = \left( \frac{d}{R} \right)^{2} f ,
        \end{equation}
where d is the distance from Earth and R is the radius of the star \citep{gray92}. 
If R is in units of solar radii and d is in pc, this relation can be written as

        \begin{equation}
         F = 1.96249 x 10^{15}\left(\frac{d}{R}\right)^{2} f .
        \end{equation}

Equation (2) can be written as
        \begin{equation}
        \frac{F}{f} = \left( \frac{d}{R} \right)^{2} = \left( \frac{2}{\Theta} \right)^{2},
        \end{equation}
where ${\rm \Theta} (= 2R / d)$ is the angular diameter of the star, in radians. For double-lined
uneclipsing binary stars, like UX Ari, ${\rm \Theta}$ could be taken as ${\rm \Theta} = 2 (R_{a} + R_{b})^2 / d$. Here R$_{a}$ and R$_{b}$ are the radius of the component stars of a binary system. Since $(R_{a} + R_{b}) \ll d$, $(R_{a} + R_{b})^2 \approx (R_{a})^2 + (R_{b})^2$. Then, Equation (2) can be written for binary system as
        \begin{equation}
        F = \left( \frac{d^2}{R_{a}^{2} + R_{b}^{2}} \right) f.
        \end{equation} 
The distance of {\rm UX Ari} is given as 50 pc by \citet{stras88} and \citet{stras93}. The recent and most reliable measurement of the distance (50.23 pc) is given in the Hipparcos Catalogue \citep{perry97}. Since the standard error of the parallax of {\rm UX Ari} is 1.25 mas, given by \citet{perry97}, the distance of 50 pc can be adopted as a good estimate. 

There are three remarkable estimations on the radii of the components of {\rm UX Ari}. Therefore, the three 
conditions could be taken into consideration for {\rm UX Ari} system:

1. With a distance of 50 pc and component radii \( R_{G5}  =  0.93 R_{\odot}\), \( R_{K0} = 3 R_{\odot}\) \citep{stras88}, Equation (3) gives:

 	\begin{equation}
         F = 4.906225 x 10^{17} f ,
        \end{equation}

   where \( R^{2} = R_{G5}^{2} + R_{K0}^{2} \)  for {\rm UX Ari} system. 

2. With the same distance as given in (1), but radii \( R_{G5} = 0.93 R_{\odot} \) and \( R_{K0} = 4.7 R_{\odot} \) \citep{stras93},  this relation becomes:

        \begin{equation}
         F = 2.137332334 x 10^{17} f .
        \end{equation}

3. With a distance of 50 pc and radii \( R_{G5} = 0.80 R_{\odot} \), \( R_{K0} = 6.2 R_{\odot} \) \citep{huen89}, this relation is

        \begin{equation}
         F = 1.255431167 x 10^{17} f .
        \end{equation}

By using Equations 4, 5 and 6 the integrated surface fluxes of the UV continuum were computed for each aspect given above, between 2100  and  3200 $\AA$.

For the comparison stars, {\rm $\eta$ Cep} and {\rm $\kappa$ Cet}, this relation can be obtained as follows.
At a distance of 14.34 pc \citep{perry97} and the radius of  \( R = 4 R_{\odot} \) \citep{black94} for {\rm $\eta$ Cep} ( K0 IV ) Equation 3 gives:

	\begin{equation}
         F = 2.522236304 x 10^{16} f ,
        \end{equation}
and, with the distance of 9.16 pc \citep{perry97} and the radius of \( R = 0.9313 R_{\odot} \) \citep{black94}
for the {\rm $\kappa$ Cet} ( G5 V ) Equation 3 gives:

	\begin{equation}
         F = 1.898537562 x 10^{17} f .
        \end{equation}
        
The integrated continuum fluxes measured on Earth and corresponding surface fluxes between 2100 and 3200 $\AA$ spectral range obtained by means of the relations given above are listed in Table 3 for {\rm UX Ari} system, and in Table 4 for {\rm $\eta$ Cep} and {\rm $\kappa$ Cet} together with IUE images. It can be seen that the fluxes obtained from LWR04857S are much lower than that those obtained from other two images. The reason is that the LWR04857S image was taken by using small aperture of the spectrograph while others were taken by large aperture. 

The effective temoeratures and the radii of comparison stars, {\rm $\eta$ Cep} and {\rm $\kappa$ Cet},
are comparible with those of the component stars of {\rm UX Ari}. Namely,\\
$\bullet$ {\rm $\eta$ Cep}(K0 IV); T$_{e}$= 4967 K \citep{soub08};\\ 
$\bullet$ {\rm UX Ari}(K0 IV); T$_{e}$= 4750 K \citep{vogt91};\\
$\bullet$ {\rm $\eta$ Cep}(K0 IV); R = 4 R$_{\odot}$ \citep{black94};\\ 
$\bullet$ {\rm UX Ari}(K0 IV); R = 4.7 R$_{\odot}$ \citep{stras93};\\
$\bullet$ {\rm $\kappa$ Cet}(G5 V); T$_{e}$= 5708 K \citep{soub08};\\
$\bullet$ {\rm UX Ari}(G5 V); T$_{e}$= 5700 K \citep{vogt91};\\
$\bullet$ {\rm $\kappa$ Cet}(G5 V); R = 0.93 R$_{\odot}$ \citep{black94};\\
$\bullet$ {\rm UX Ari}(G5 V); R = 0.93 R$_{\odot}$ \citep{stras93};\\
Therefore, recalling that the sum of the fluxes of two components can be used in computing the magnitude
of a binary sysytem, like {\rm UX Ari}, by\\
\begin{equation}
m_{1} - m_{s} = -2.5 log \left(\frac{f_{1}}{f_{1}+f_{2}}\right),
\end{equation}  
where the subscription '1' and '2' are refer to the component stars, and 's' refers to the system, the theoretical continuum flux of the {\rm UX Ari} system can be estimated by adjusting for its distance of 50 pc and using the observed continuum fluxes of {\rm $\eta$ Cep}(K0 IV) and {\rm $\kappa$ Cet}(G5 V):

	\begin{equation}
         f_{theo} (Ari) = f_{\eta  Cep} + f_{\kappa  Cet} .
        \end{equation}

        \begin{table*}[h!p]
        \scriptsize
        \caption{Integrated continuum fluxes}
        \medskip
        \begin{center}
        \begin{tabular}{lccccccc}
        \hline
        Image & HJD(mid)& Epoch & f & UV excess & F$_{Str88}$& F$_{Huen89}$ & F$_{Str93}$ \\
              &         &       &(x10$^{-10}$)& in $f(\%)$&($10^{7} erg cm^{-2} s^{-1}$)&($10^{7} erg cm^{-2} s^{-1}$)& ($10^{7} erg cm^{-2} s^{-1}$) \\
        \hline
        LWR06329S & 2444215.975 &  634.089 & 1.44(0.06)& - & 7.07(0.28)& 1.81(0.07)& 3.08(0.12)\\
        LWR06329L & 2444215.978 &  634.090 & 2.57(0.05)&  4.3 & 12.60(0.24)& 3.22(0.06)& 5.49(0.11) \\
        LWP11747  & 2447067.937 & 1077.084 & 2.97(0.05)& 20.6 & 14.57(0.26)& 3.73(0.07)& 6.35(0.11) \\
        LWP11748  & 2447067.960 & 1077.088 & 3.05(0.05)& 24.0 & 14.98(0.26)& 3.83(0.07)& 6.53(0.12) \\
        LWP11751  & 2447068.238 & 1077.131 & 2.91(0.07)& 18.4 & 14.30(0.32)& 3.66(0.08)& 6.23(0.14) \\
        LWP11752  & 2447068.364 & 1077.150 & 2.89(0.06)& 17.4 & 14.18(0.27)& 3.63(0.07)& 6.18(0.12) \\
        LWP11753  & 2447068.484 & 1077.169 & 2.83(0.05)& 15.1 & 13.90(0.27)& 3.56(0.07)& 6.06(0.12) \\
        LWP11754  & 2447068.612 & 1077.189 & 2.84(0.05)& 15.3 & 13.92(0.27)& 3.56(0.07)& 6.07(0.12) \\
        LWP11755  & 2447068.732 & 1077.208 & 2.79(0.06)& 13.5 & 13.71(0.27)& 3.51(0.07)& 5.97(0.12) \\
        LWP11760  & 2447068.973 & 1077.245 & 2.81(0.06)& 14.2 & 13.79(0.27)& 3.53(0.07)& 6.01(0.12) \\
        LWP11761  & 2447068.998 & 1077.249 & 2.86(0.05)& 16.0 & 14.01(0.24)& 3.59(0.06)& 6.10(0.11) \\
        LWP11763  & 2447069.085 & 1077.262 & 2.86(0.06)& 16.0 & 14.02(0.27)& 3.59(0.07)& 6.11(0.12) \\
	LWP11764  & 2447069.234 & 1077.286 & 2.88(0.05)& 16.9 & 14.12(0.27)& 3.61(0.07)& 6.15(0.12) \\
	LWP11765  & 2447069.360 & 1077.305 & 2.92(0.06)& 18.6 & 14.33(0.27)& 3.67(0.07)& 6.24(0.12) \\
	LWP11766  & 2447069.483 & 1077.324 & 2.84(0.05)& 15.3 & 13.92(0.27)& 3.56(0.07)& 6.07(0.12) \\
        LWP11767  & 2447069.588 & 1077.341 & 2.93(0.06)& 19.1 & 14.39(0.28)& 3.68(0.07)& 6.27(0.12) \\
        LWP11768  & 2447069.614 & 1077.345 & 2.93(0.06)& 19.1 & 14.38(0.28)& 3.68(0.07)& 6.27(0.12) \\
        LWP11769  & 2447069.638 & 1077.348 & 2.87(0.06)& 16.5 & 14.08(0.28)& 3.60(0.07)& 6.13(0.12) \\
        LWP11770  & 2447069.664 & 1077.352 & 2.95(0.06)& 19.7 & 14.46(0.29)& 3.70(0.07)& 6.30(0.13) \\
	LWP31888  & 2450100.379 & 1548.113 & 2.51(0.05)&  1.8 & 12.29(0.23)& 3.15(0.06)& 5.35(0.10) \\
	LWP31894  & 2450103.355 & 1548.575 & 2.32(0.05)& - & 11.39(0.23)& 2.92(0.06)& 4.96(0.10) \\
	LWP31895  & 2450103.389 & 1548.581 & 2.39(0.05)& - & 11.72(0.23)& 3.00(0.06)& 5.11(0.10) \\
	LWP31896  & 2450103.425 & 1548.586 & 2.48(0.05)&  1.0 & 12.17(0.24)& 3.11(0.06)& 5.30(0.10) \\
	LWP31903  & 2450105.400 & 1548.893 & 2.55(0.05)&  3.4 & 12.49(0.23)& 3.20(0.06)& 5.44(0.10) \\
        \hline
        \end{tabular}
        \end{center}
        \end{table*}

        \begin{table*}[h!p]
        \small
        \caption{Integrated continuum fluxes of {\rm $\eta$ Cep} and {\rm $\kappa$ Cet}}
        \medskip
        \begin{center}
        \begin{tabular}{lccccc}
        \hline
        Image & HJD(mid)& Star & f & F & f$_{50}$ \\
              &         &      &($10^{-10}$ erg $cm^{-2}$ $s^{-1}$)&($10^{7}$ erg $cm^{-2}$ $s^{-1}$)&($10^{-10}$ erg $cm^{-2}$ $s^{-1}$)\\
        \hline
        LWR12739  & 2445037.093 & $\eta$ Cep & 23.32(0.18)& 5.88(0.04)& 1.92(0.01)\\
        \\
        LWR04857S & 2444048.950 & $\kappa$ Cet & 10.56(0.14)& 20.06(0.26)& 0.35(0.05)\\
        LWR04857L & 2444048.955 & $\kappa$ Cet & 17.13(0.16)& 32.52(0.29)& 0.57(0.05)\\
        LWR04858  & 2444048.982 & $\kappa$ Cet & 15.29(0.82)& 29.03(0.16)& 0.51(0.03)\\
	    \hline
        \end{tabular}
        \end{center}
        \end{table*}

This theoretical flux can be used for the examination of ultraviolet excess of {\rm UX Ari} by comparison with the observed continuum fluxes. Here, the theoretical surface fluxes of the comparison stars (see Column 5 of Table 4), {\rm $\kappa$ Cet} and {\rm $\eta$ Cep}, were used for evaluating the observed fluxes with adjustment made for the distance of UX Ari. The comparison of observed fluxes (see Column 4 of Table 3) with those of 50 pc observed fluxes (adjustment to the distance of {\rm UX Ari}, see the last column of Table 4) of {\rm $\kappa$ Cet} and {\rm $\eta$ Cep} show that there is some UV excess in UX Ari, which varies from 1\% up to 24\%, with the exception of two images (namely LWP31894 and LWP31895). In this evaluation LWR06329S image was excluded because it was taken with small aperture of the spectrograph. The adjustment 50 pc - observed 
fluxes of comparison stars were computed using the corresponding surface fluxes given in the Column 5 of Table 4. 

\subsection{Another Test of the UV Excess in {\rm UX Ari}}
 
Although the surface fluxes are the most important data in this spectral analysis, it is also useful to look at the  result of the testing of the UV excess by taking only the observed fluxes at Earth into consideration, as in the spectrophotometric analysis of  \citet{rhomb77}. 

By using Equation (12) written as

\begin{equation}
         f_{theo} (Ari) = C_{1} f_{\eta  Cep} + C_{2} f_{\kappa  Cet}  ,
        \end{equation}

where 

\begin{figure*}[h!t]
\vspace{0.1cm}
\begin{center}
\includegraphics[angle=0, width=0.75\textwidth]{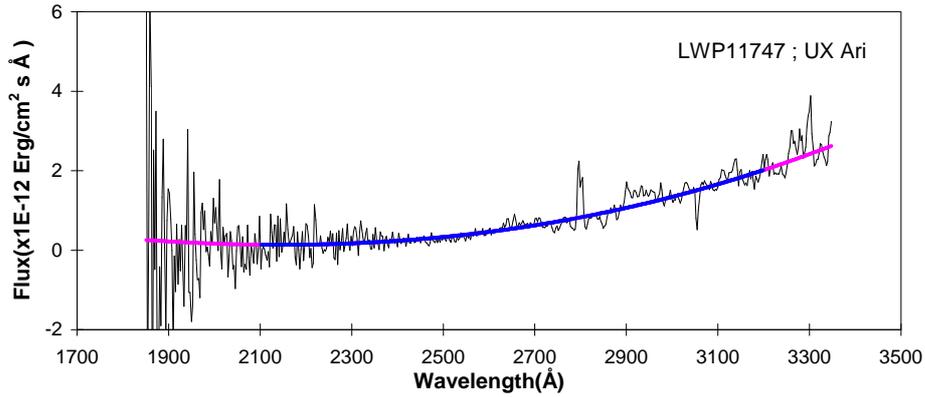}
\caption{A sample of the third-order polynomial fitting to the UV continuum of the long-wavelength, low-dispersion IUE spectrum, LWP11747.}
\label{fig5}
\end{center}
\end{figure*}

\begin{equation}
C_{1} = \left( \frac {R_{K0 UXAri}} {R_{Eta Cep}} \right)^{2} \left( \frac {d_{Eta Cep}} {d_{UX Ari}} \right)^{2} = 0.11356226,
\end{equation}      
  with R$_{K0 UXAri}$= 4.7 R$_{\odot}$ and
\begin{equation}        
C_{2} =  \left( \frac{R_{G5 UXAri}} {R_{Kap Cet}} \right)^{2} \left( \frac {d_{Kap Cet}}{d_{UX Ari}} \right)^{2} = 0.033468606,
\end{equation}         
with R$_{K0 UXAri}$= 0.93 R$_{\odot}$.        
Using the values of integrated fluxes measured at Earth of $1.92\times 10^{-10}$ for {\rm $\eta$ Cep} and
of $0.54\times 10^{-10}$ (mean value) for {\rm $\kappa$ Cet} (see Table 4), we have 
\begin{equation}
f_{theo} (Ari) = f_{Eta Cep} + f_{Kap Cet} = 2.46\times 10^{-10}.
\end{equation}
By examination of the ratio $f_{UX Ari}$ / $f_{theo}$ (Ari) = 1.01 to 1.24. Depending on wavelength, the calculation using the observed UV fluxes at Earth in Equation 12 (second approach) shows also that there is some ultraviolet excess in the {\rm UX Ari} system.

\begin{figure*}[h!t]
\vspace{0.1cm}
\begin{center}
\includegraphics[angle=0, width=0.75\textwidth]{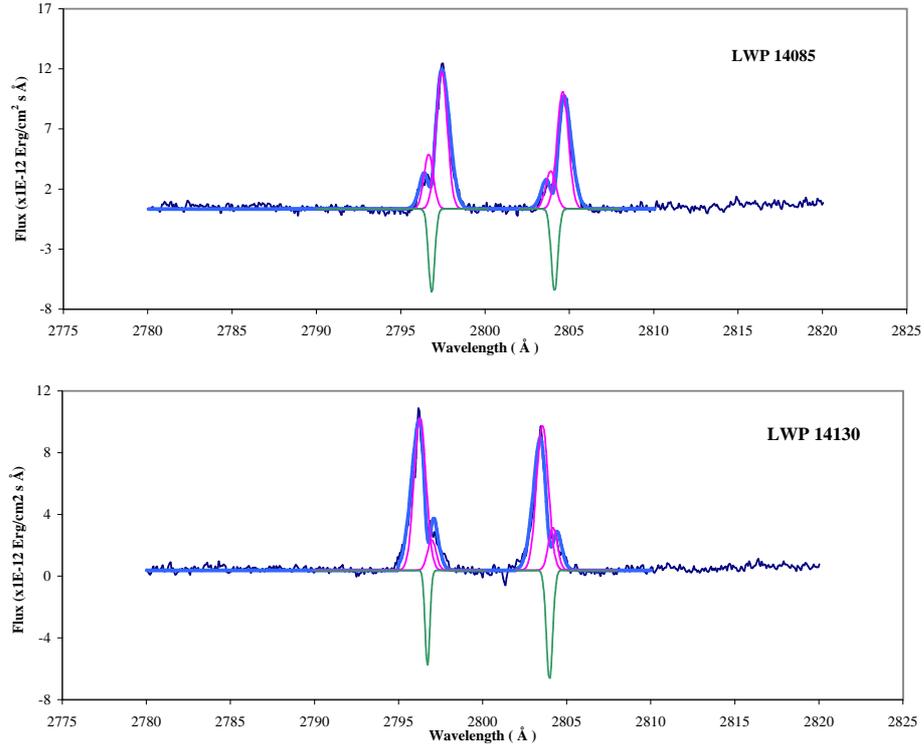}
\caption{The Gaussian fits to MgII h and k lines of double-lined spectroscopic binary {\rm UX Ari} appeared in LWP14085 and LWP14130 high resolution images. The explanation for all components of Gasussian profiles are given in Section 2.}
\label{fig6}
\end{center}
\end{figure*}

\subsection{Long-Wavelength, High - Dispersion Spectra}

The most prominent features seen in these spectra are the well-known chromospheric Mg II h and k emission lines. In all LWP and LWR high resolution images, these Mg II h and k line profiles of both K0 IV and G5 V stars, appeared to be compatible with the corresponding orbital phases (Figures 6 and 7). Therefore, the Mg II h and k flux variation can also be evaluated depending on orbital phase. Based on the fitting procedure mentioned at the beginning of Section 2, the integrated line fluxes, the equivalent widths and the radial velocities for all components of the Mg II profiles were computed. The integrated line fluxes of Mg II k and MgII h and k, from G5 V and K0 IV component, are plotted in Figures 8 and 9 as a function of time (in the sense of epoch) and orbital phase, respectively. Similar trends were seen for the total Mg II line fluxes of both components (G5 V + K0 IV) of {\rm UX Ari}. The scattering of Mg II line fluxes that appeared in Figure 8 
is similar to that in Figures 2-4. This scattering was also attributed to the flux variation (which showed the maxima near 0.20 and 0.70 orbital phases) with the orbital phase. The Mg II h and k radial velocity curves of {\rm UX Ari} system are shown in Figure 10 together with the results of \citet{duem01} and of \citet{carlos71} obtained from the visible spectral range. It is seen that the velocities of K0 IV component are in a better aggrement with the optical data than the velocities of G5 V component. Recalling that the effect of interstellar absorption has been removed by the Gaussian profile fitting procedures (see Section 2), this greater scattering in the G5 V velocities is likely due to the physical interaction between the K0 IV and G5 V compenent of {\rm UX Ari}, just as the mass exchange via coronal/magnetic loops. The velocity $\gamma$  of the centre of mass of the system was found to be $38.22 \pm 2.36$ {\kms} from the sinusoidal fitting to the Mg II h radial velocity curve  and $31.01 \pm 2.44$ {\kms} from the same analysis of the Mg II k radial velocity curve. Therefore, the mean value of  $\gamma$ is found to be $34.62 \pm 4.86$ {\kms} from the Mg II h and k radial velocity curves of {\rm UX Ari}. This $\gamma$ is somewhat greater (about 8 {\kms}) than that of 
\citet{duem01}'s and \citet{carlos71}'s value (26.5 {\kms}).

\section{Discussion and Conclusions}

The conclusions of this study together with related discussion are as follows:
 
All integrated emission line fluxes of short wavelength low dispersion spectra showed a clear variation with time and orbital phase, but the variation with time was not as clear as that with the orbital phase (Figures 2 to 4). For example, the spectra taken in 1978(near the epoch of 560) showed a bit larger scattering in the range of fluxes. But when plotting these data versus orbital phase, the flux distribution showed a more clear flux variation. This feature seems to be similar in the other epochs (epochs of 963, 1035, 1301 and 1468). Apart from the flare event shown on 1979 Jan 1 (SWP03766), there were some flux enhancements (especially in the lines originating in the middle and upper chromosphere) in the years of 1987 (epoch of 1035.732), 1991 (epoch of 1300.757) and 1994 (near the epoch of 1470). However, the periodicity of the flux variation in time was not detected clearly from the 18 years of data. An application of the period search by discrete Fourier 
transform to OI (middle chromosphere line), CII (upper chromosphere line), SiIV (transition region line), and MgII k emission line fluxes did not give the significant results due to large gaps in the IUE data. On the other hand, the evaluation of the highest flux level of emission lines (occured at some epochs) showed that the first enhancement was in 1987 (epoch of 1035) occurring 9 years after the first data was obtained in 1978. If the period was 9 years, the next flux enhancement should have been in 1996 (epoch of 1548) instead of 1994 (epoch of 1468). Therefore, the variation with time may be with the periodicity of 7 - 9 years \citep[which is close to the well known 10 years cycle of the {\rm RS CVn} phenomena, see][]{rodono80}. However, this variation was not as clear as that with orbital phase, probably due to the insufficient distribution of the data to determine such a periodicity.

\begin{figure*}[h!t]
\vspace{0.1cm}
\begin{center}
\includegraphics[angle=0, width=0.65\textwidth]{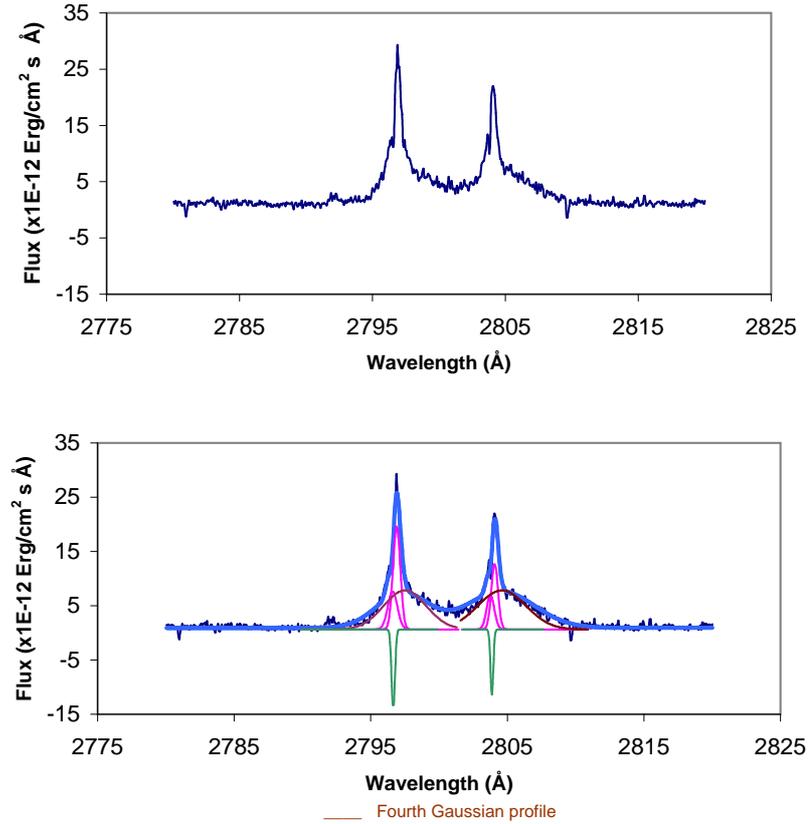}
\caption{The IUE flare spectrum of {\rm UX Ari} observed on 1979 Jan 1 at 0.062 orbital phase. In the Gaussian profiles fitting procedure the fourth component was added for the flare effect on the MgII h and k profiles.}
\label{fig7}
\end{center}
\end{figure*}
 
There were two explicit increments around 0.20P and 0.70P in all chromospheric and transition line flux variations depending on the orbital phase. The first flux increment near the 0.20P was composed of the data from the spectra taken in 1981 (SWP13162), in 1990 (SWP39460) and in 1994 (SWP51866 , SWP51867, SWP51961, SWP51962, SWP52016, SWP52017). The second flux increment near the 0.70P was composed of the data from the spectra taken in 1979 (SWP07267), in 1987(SWP30026, SWP30027, SWP30028, SWP30029, SWP30030), in 1991 (SWP42416) and in 1994 (SWP52046, SWP52047). The {\it V} light-curve amplitudes of {\rm UX Ari} in these years are 0.16 mag (1981), 0.07 mag (1990), 0.19 mag (1994), 0.04 mag (1979), 0.19 mag (1987) and 0.06 mag (1991) \citep{rave95}. Therefore, these flux increments did not seem to correlate with the {\it V} light-curve amplitudes, but there was a good agreement with the configuration of the component stars near the quadratures. The same situation also appeared in the MgII h and k emission-line fluxes. By using several optical chromospheric activity indicators, HeI D$_{3}$, NaI D$_{1}$, D$_{2}$, H$_{\alpha}$ and CaII IRT lines \citet{gu02} also detected this high activity level of {\rm UX Ari} around the second quadrature. They suggested that this may originate in the coupling of the chromospheric activity of the secondary and mass-transfer activity of the two components. Another important consideration is that HeII ($\lambda$1640) fluxes may contribute to flux enhancement due to collisional excitation (Athay  1965 ; Jordan  1975) indicating a temperature of  $\sim$ 8 x $10^{4}$ K and recombination following photoionization by coronal X-rays \citep{zirin76}. The contribution of recombination to the HeII flux increases it up to 80\%  in the more 
active region \citep{rego83}. Another contributor to HeII is the FeII $\lambda$1640.15 emission (Jordan, 1975; Kohl, 1977). Therefore, the HeII ($\lambda$1640) emission feature cannot be considered a pure chromospheric indicator for {\rm UX Ari}.
 
Examination of the ultraviolet excess in {\rm UX Ari} by using the 24 long-wavelength, low-dispersion spectra of {\rm UX Ari} (see Table 3), and comparison stars {\rm $\kappa$ Cet} and {\rm $\eta$ Cep} (see Table 4) in the spectral range between $2100 \AA$  and $3200 \AA$ showed that there is some ultraviolet excess in {\rm UX Ari} system which varies from 1\% up to 24\%. However, two of these images, LWP31894 and LWP31895, showed no ultraviolet excess for {\rm UX Ari} system. These 24 spectra were taken in the 1979-1996 period and covered most of the orbital phases. This examination is based on the computation of the theoretical continuum surface fluxes (computed from {\rm $\kappa$ Cet} and {\rm $\eta$ Cep} spectra as comparison stars) and the {\rm UX Ari} continuum surface fluxes in the spectral range mentioned above. Using the same comparison stars ({\rm $\kappa$ Cet} and {\rm $\eta$ Cep}) and based on their 1975 observations, \citet{rhomb77} measured spectrophotometrically the ultraviolet excess in {\rm UX Ari} in which the cool star contributes $75\% \pm 5\%$ of the total light of the system at $4700 \AA$. Their spectrophotometric observations were carried out at orbital phases 0.715 and 0.791, and they attributed the wavelength-dependent ultraviolet excess in {\rm UX Ari} to the free-free emission from hot circumstellar gas in the system. There is clearly agreement with the results of \citet{rhomb77} and this study on  the  existence of ultraviolet excess in {\rm UX Ari}.

\begin{figure}[h!t]
\vspace{0.1cm}
\begin{center}
\includegraphics[width=7.5cm]{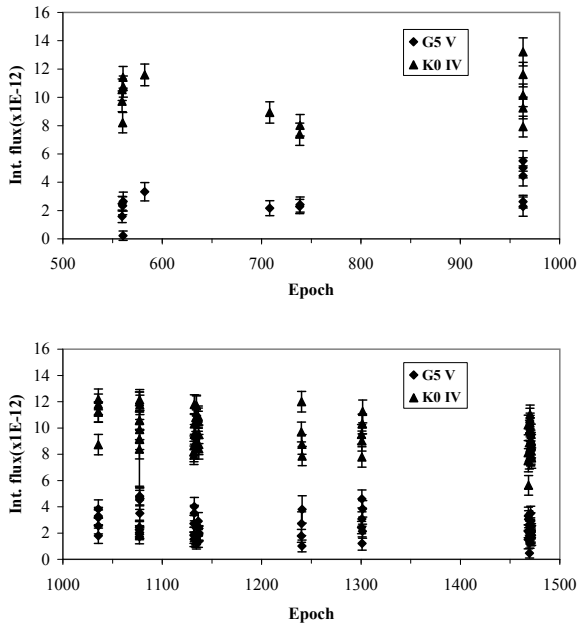}
\caption{The integrated MgII k line fluxes of the components of {\rm UX Ari} system as a function of time. The fluxes are in units of erg cm-2 s-1.}
\label{fig8}
\end{center}
\end{figure}

\begin{figure}[h!]
\vspace{0.1cm}
\begin{center}
\includegraphics[width=7.5cm]{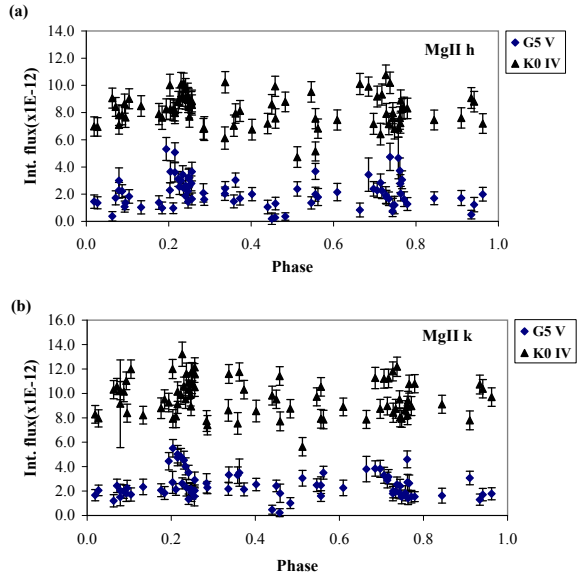}
\caption{(a) The MgII h line fluxes, and (b) the MgII k line fluxes of the components of {\rm UX Ari} system as a function of orbital phase. The fluxes are in units of erg cm-2 s-1.}
\label{fig9}
\end{center}
\end{figure}
 
The integrated UV continuum fluxes of {\rm UX Ari} have the lowest flux level of long-wavelength, low-dispersion IUE spectra near the epoch of 634 and these two images (LWR06329L and LWR06329S) were taken about eleven months after the flare event that occured on 1979 Jan 1. At the time of LWR06329 images (1979 Dec. 8), the {\it V} light curve of {\rm UX Ari} had an amplitude about 0.04 mag \citep{rave95}. The integrated UV continuum fluxes of {\rm UX Ari}, near the epoch of 1077 (see Table 3), show the variaton with the orbital phase, and have the highest flux level of long-wavelength low dispersion IUE spectra. After the flare event appeared in January of 1987, detected from simultaneous IUE and VLA observations \citep{lang88}, during the time interval from 1987 September 29  to  1987 October 1 (the dates of images from LWP11747 to LWP11770 ; see Table 3), the {\it V} light curve of {\rm UX Ari} had an amplitude of about 0.19 mag \citep{rave95}. Near the epoch of 1548 (1996 January), the integrated UV continuum fluxes of {\rm UX Ari} (see Table 3) show the variation with the orbital phase and have lower flux levels than that of the IUE spectra taken in 1987 (near the epoch of 1077). Especially the three images, taken sequentially (LWP31894, LWP31895  and LWP31896) on 1996 Jan 20 show some rise in the UV continuum fluxes near the 0.6 orbital phase. Unfortunately, no photometric, X-ray or radio observation of {\rm UX Ari} made at this time interval of epoch 1548 has been made to enable comparison with these UV continuum fluxes. 

\begin{figure}[h!t]
\vspace{0.1cm}
\begin{center}
\includegraphics[width=7.5cm]{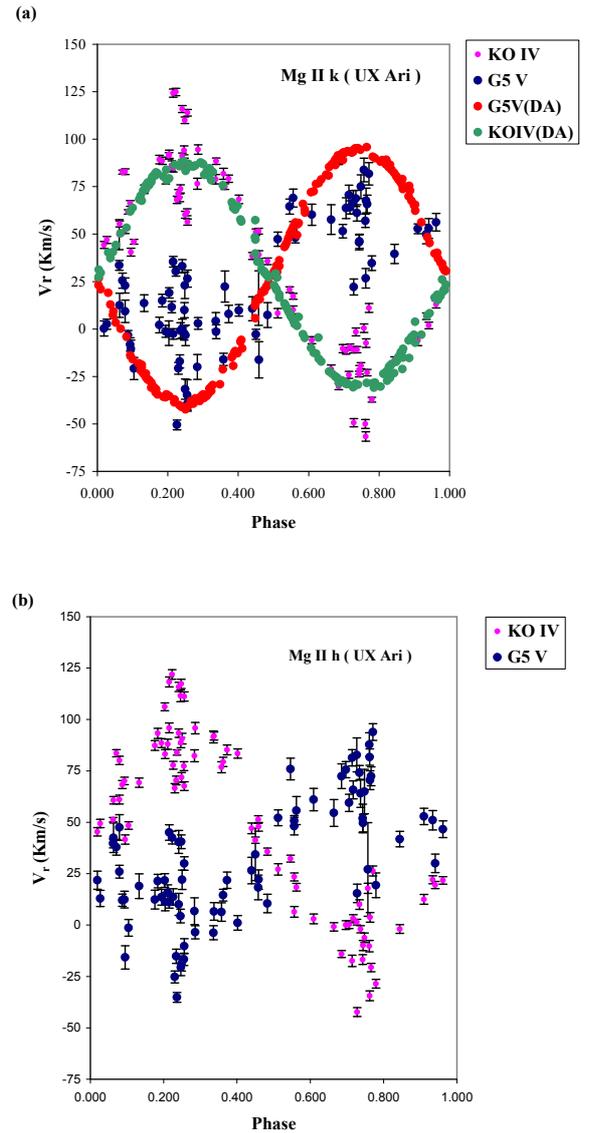}
\caption{(a) The MgII k radial velocity curves of {\rm UX Ari} system together with the curves of \citet{duem01} (with the (DA) legends) which listed in their Table 2, and (b) the MgII h radial
velocity curves of {\rm UX Ari}.}
\label{fig10}
\end{center}
\end{figure}
 
Similar to the emission line fluxes of low dispersion spectra, MgII h and k emission line fluxes of long wavelength high resolution spectra (evaluated from 79 images) have a variation with the orbital phase (Figure 9). There are also some increments of fluxes near 0.20 and 0.70 orbital phases. From the individual MgII emission-line fluxes of the component stars of {\rm UX Ari} (see Figure 9), it was found that the contributions to the activity of the system of G5 V and K0 IV components are about, on average, 20\% and 80\% respectively, but these ratios varied with time and the orbital phase. However, as mentioned in the introduction of this paper, these contributions are estimated to be about 25\% for G5 V and 75\% for K0 IV from 25 images with IUESIPS reduction, which are not much different from the evaluation given above. Therefore, it can be pointed out that the activity of the system not only comes from the K0 IV component but also comes partially from G5 V component of {\rm UX Ari}. That means both components have activity phenomena with the most of the contributions to the activity of the system coming from K0 IV. As direct evidence for the 
activity level of the secondary star of {\rm UX Ari}, this result confirms the findings and discussions on the secondary component given by \citet{aarum3}, by using CaII K core emission of the secondary.
 
Although the absorption feature, observed on the peak of the MgII k profiles of the K0 IV component and shifted together with the emission profile as the star revolving on its orbit, was determined on the IUE spectra with IUESIPS reduction \citep{ekm93} this absorption feature was not found to appear on the IUE spectra with NEWSIPS reduction (Figure 6). This discrepancy is likely to originate from the absolute flux errors in the IUESIPS processing summarized by \citet{nichols96}.
 
There is a flare event observed on 1979 Jan 1 at 0.062 orbital phase (LWR 03344). Also, this flare event appeared in the short wavelength low dispersion IUE image SWP 03766 taken on the same date at 0.068 orbital phase. These two images were also studied by \citet{simon80} who gave a plausible explanation for this flare emission, stating that the downflowing material from K0 IV component onto the G5 V star with velocities ranging up to 475 km $s^{-1}$  possibly originates in stellar prominences, or at the base of coronal loops associated with the active regions on the surface of the K0 IV star, or with material streaming between the stars. Their flux estimations for this flare spectrum were 3.8 x $10^{-11}$  ergs $cm^{-2}$ $s^{-1}$  for MgII k  and  3.2 x $10^{-11}$ ergs $cm^{-2}$ $s^{-1}$  for MgII h emission which are slightly different from the values estimated in this study ( 4.8 x $10^{-11}$ ergs $cm^{-2}$ $s^{-1}$ for MgII k  and 4.7 x $10^{-11}$ ergs $cm^{-2}$ $s^{-1}$  for MgII h ). These differences could have mainly arisen from different reduction procedures(IUESIPS/NEWSIPS). In the Gaussian profile fitting procedure for this flare spectrum (LWR03344) a fourth Gaussian profile (denoted by brown solid line in Figure 7) was added to match the flare event for both h and k emission lines. The flare contribution to the integrated emission line fluxes not only comes from this fourth Gaussian component but also from the G5 V and K0 IV components. The total effect of this flare must be shared, with appropriate amounts, between the G5 V, K0 IV and the conjunction of coronal loops between the component stars of {\rm UX Ari} system which is located nearer to G5 V star \citep[see Figure 4 of][]{simon80}.
 
Comparison of the radial velocities of MgII k emission line profiles of the components of {\rm UX Ari} system with those of radial velocities of \citet{duem01} and of \citet{carlos71} obtained from the visible spectral range showed some agreements with the data of K0 IV component but the velocities of MgII for G5 V component were, mostly near the quadratures, much different (lower) than the velocities obtained from the visible spectral range (Figure 10a). In addition, the velocities of MgII for G5 V component showed a great scattering by comparison with velocities of MgII for K0 IV component. On the account of the effect of the interstellar absorption was removed by the Gaussian profile fitting procedure, this great scattering in the velocities of G5 V component could likely due to physical interaction between the K0 IV and G5 V component which is seen actively in UV spectral region. This great scattering and the lower velocities compared to visible data, could make a suggestion for the chromospheric activity via a magnetic dynamo which produced the active region loops. 
Moreover, the chromospheric instability of G5 V could be due to interaction between the G5 V and K0 IV components  via magnetic coronal loops. The mean value of the velocity $\gamma$  of the centre of mass of the system was found to be $34.62 \pm 4.86$ {\kms} from the MgII h and k radial-velocity curves, which was somewhat greater(about 8 {\kms}) than that of \citet{duem01} and \citet{carlos71} (26.5 {\kms}).\\

Combining all spectral characteristics of {\rm UX Ari} supported the model of inhomogeneous gyro-synchrotron 
emission arising from electrons which have interaction with inhomogeneous magnetic fields \citep{mutel87}. 
As a result of these agreements, the UV emission flux variation with orbital phase in {\rm UX Ari} might be 
strongly correlated with the size and configuration of dark spots (see  Vogt \& Hatzes 1991  and  Aarum \& Engvold 2003) related to the magnetic origin of the activity phenomena. Based on the contribution of G5 V and K0 IV components (on the order of 20\%  and  80\% , respectively) to the MgII activity of the system, it would be suggested that it is better to take into consideration the spot distribution not only on the surface of K0 IV  but also on the surface of the G5 V component of {\rm UX Ari}. However, some constraints on the secondary component were given by \citet{aarum3}.

\section*{Acknowledgments}
I would like to thank Randy Thompson for his kind help in converting NEWSIPS files to ASCII formats by 
using IDL on the IUE account. I also thank Mesut Y{\i}lmaz and Tolga \c{C}olak for their assistance in 
compiling the manuscript with latex. And finally, I would like to thank the referee for his / her cautions 
on some points to improve the result of this study. This research has made use of the Simbad database, 
operated at CDS, Strasbourg, France, and of NASA's Astrophsics Data System Bibliographic Services.


\begin{thebibliography}{99}

\bibitem[Aarum Ulv\aa s \& Engvold(2003)]{aarum3} Aarum Ulv\aa s, V.,  Engvold, O., 2003, {\rm A\&A} 402, 1043   
\bibitem[Athay(1965)]{athay65} Athay, R.G., 1965, {\rm ApJ}, 142, 755
\bibitem[Bevington(1969)]{beving69} Bevington, P. R., 1969, Data Reduction and Error Analysis for 
                             The Physical Sciences, Mc Graw Hill, New York, p. 237
\bibitem[Blackwell \& Lynas-Gray(1994)]{black94} Blackwell, D.E., Lynas-Gray, A.E., 1994, {\rm A\&A}, 282, 899
\bibitem[Carlos \& Popper(1971)]{carlos71} Carlos, R.C., Popper, D.M., 1971, {\rm PASP}, 83, 504
\bibitem[Duemmler \& Aarum(2001)]{duem01} Duemmler, R., Aarum, V., 2001, {\rm A\&A}, 370, 974 
\bibitem[Ekmek\c{c}i(1993)]{ekm93} Ekmek\c{c}i, F., 1993, PhD Thesis, Ankara Uni. Graduate School of Natural and Applied
                              Sciences, Dept. of Astronomy and Space Sciences       
\bibitem[Gray(1992)]{gray92} Gray, D. F., 1992, The Observations and Analysis of Stellar
                          Photospheres, Second Edition, Cambridge Univ. Press , New York, p. 340
\bibitem[Gu et al.(2002)]{gu02} Gu, S.-h., Tan, H.-s., Shan, H.-g., Zhang, F.-h., 2002, {\rm A\&A}, 388, 889                                                       
\bibitem[Huenemoerder, Buzasi \& Ramsey(1989)]{huen89} Huenemoerder, D. P., Buzasi, D. L., Ramsey, L. W., 1989, {\rm AJ}, 98, 1398  
\bibitem[Jordan(1975)]{jordan75} Jordan, C., 1975, {\rm MNRAS}, 170, 429    
\bibitem[Kohl(1977)]{kohl77} Kohl, J.L., 1977, {\rm ApJ}, 211, 958    
\bibitem[Lang \& Willson(1988)]{lang88} Lang, K.R.,  Willson, R.F., 1988, {\rm ApJ}, 328, 610
\bibitem[Montes et al.(1995a)]{montes95a} Montes, D., Fern\'{a}ndez-Figueroa, M. J., De Castro, E., Cornide, M., 1995a, {\rm A\&AS}, 109, 135
\bibitem[Montes et al.(1995b)]{montes95b} Montes, D., Fern\'{a}ndez-Figueroa, M. J., De Castro, E., Cornide, M., 1995b, {\rm A\&A}, 294, 165        
\bibitem[Mutel et al.(1987)]{mutel87} Mutel, R.L., Morris, D.H., Doiron, D.J., Lestrade, J.F., 1987, {\rm AJ}, 93, 1220  
\bibitem[Nichols \& Linsky(1996)]{nichols96} Nichols, J. S., Linsky, J. L., 1996, {\rm AJ}, 111, 517
\bibitem[Perryman et al.(1997)]{perry97} Perryman, M. A. C., Lindegren, L., Kovalevsky, J., et al., 1997, {\rm A\&A}, 323L, 49        
\bibitem[Raveendran \& Mohin(1995)]{rave95} Raveendran, A. V., Mohin, S., 1995, {\rm A\&A}, 301, 788
\bibitem[Rego, Gonzales-Rieastra \& Fern\'{a}ndez-Figueroa 1983]{rego83} Rego, M., Gonzalez-Riestra, R., Fern\'{a}ndez-Figueroa, M.J., 1983, {\rm A\&A}, 119, 227    
\bibitem[Rhombs \& Fix(1977)]{rhomb77} Rhombs, C. G., Fix, J. D., 1977, {\rm ApJ}, 216, 503 
\bibitem[Rodon\'{o}(1980)]{rodono80} Rodon\'{o}, M., 1980, {\rm MmSAI}, 51, 623                                                 
\bibitem[Simon, Linsk \& Schiffer(1980)]{simon80} Simon, T., Linsky, J.L., Schiffer, F. H., 1980, {\rm ApJ}, 239, 911    
\bibitem[Soubrian et al.(2008)]{soub08} Soubiran, C., Bienaym\'{e}, O., Mishenina, T.V., \& Kovtyukh, V.V., 2008, {\rm A\&A}, 480, 91 
\bibitem[Strassmeier et al.(1988)]{stras88} Strassmeier, K. G., Hall, D. S., Zeilik, M., et al., 1988, {\rm A\&AS}, 72, 291
\bibitem[Strassmeier et al.(1993)]{stras93} Strassmeir, K. G., Hall, D. S., Fekel, F. C., Scheck, M., 1993, {\rm A\&AS}, 100, 173     
\bibitem[Vogt \& Hatzes(1991)]{vogt91} Vogt, S. S., Hatzes, A. P., 1991, in {\rm IAU} Coll. No 130, The Sun and Cool Stars. Activity,
                               Magnetism, Dynamos, Eds.Tuominen, I., Moss, D., Rudiger, G., Springer Berlin Heidelberg New York, p. 297
\bibitem[Zirin(1976)]{zirin76} Zirin, H., 1976, {\rm ApJ}, 208, 414

\end{thebibliography}
\end{document}